\documentclass[]{spie}  

\usepackage{amsmath,amsfonts,amssymb}
\usepackage{graphicx}
\usepackage[colorlinks=true, allcolors=blue]{hyperref}
\usepackage{siunitx}
\usepackage{comment}
\usepackage{float}
\usepackage{array}
\usepackage{subcaption}
\usepackage[style=base]{caption}

\title{Advancing Control Electronics for Next-Generation Astronomical Fiber Robotic Positioners}

\author[a]{Sébastien Pernecker}
\author[a]{Jonathan Wei}
\author[a]{Maxime Rombach}
\author[a]{Oliver Pineda Suárez}
\author[a]{Tarik Ibrahimovic}
\author[a]{Jean-Paul Kneib}

\affil[a]{Institute of Physics, Laboratory of Astrophysics,  École Polytechnique Fédérale de Lausanne (EPFL), Observatoire de Sauverny, CH-1290 Versoix, Switzerland}

\authorinfo{Further author information: (Send correspondence to S. P.)\\S. P.: E-mail: sebastien.pernecker@epfl.ch, Telephone: +41 21 693 97 24}

\begin{document} 
\maketitle

\begin{abstract}
Next-generation spectroscopic surveys require compact, high-density fiber robotic positioner systems achieving 5um precision, placing strict constraints on the size and the power budget of the control electronics. We present a compact control electronics architecture that drives 21 theta–phi SCARA positioners (42 BLDC motors) on a single board, representing a significant increase in complexity compared to the electronics used in ongoing surveys such as SDSS-V and DESI, where each positioner relies on dedicated hardware. The design integrates power distribution, CAN communication, and a synchronization line for simultaneous motion in high-density environments. Sensorless Field-Oriented Control with collision detection and mechanical/magnetic hard-stop calibration enables accurate positioning without Hall sensors or encoders, reducing system cost and complexity. We describe the system architecture and performance validation, demonstrating that the module meets precision requirements while reducing the space required for control electronics and improving energy efficiency for future survey instruments.
\end{abstract}


\section{Introduction}
\label{chap:introduction}

\subsection{Context and motivation}
\label{sec:context}

Ground-based astronomical surveys aim to characterize the universe at cosmological scales, advancing our understanding of dark energy, dark matter, and large-scale structure. A cornerstone observational technique in this effort is multi-object spectroscopy (MOS), in which light from a large number of astronomical targets is collected simultaneously and dispersed into spectra for analysis \cite{Hill1980MedusaMOS, Smee_2013}. The Sloan Digital Sky Survey (SDSS) is a prominent example of such a ground-based imaging and spectroscopic program.

\begin{figure}[H]
    \centering
    \includegraphics[width=0.6\textwidth]{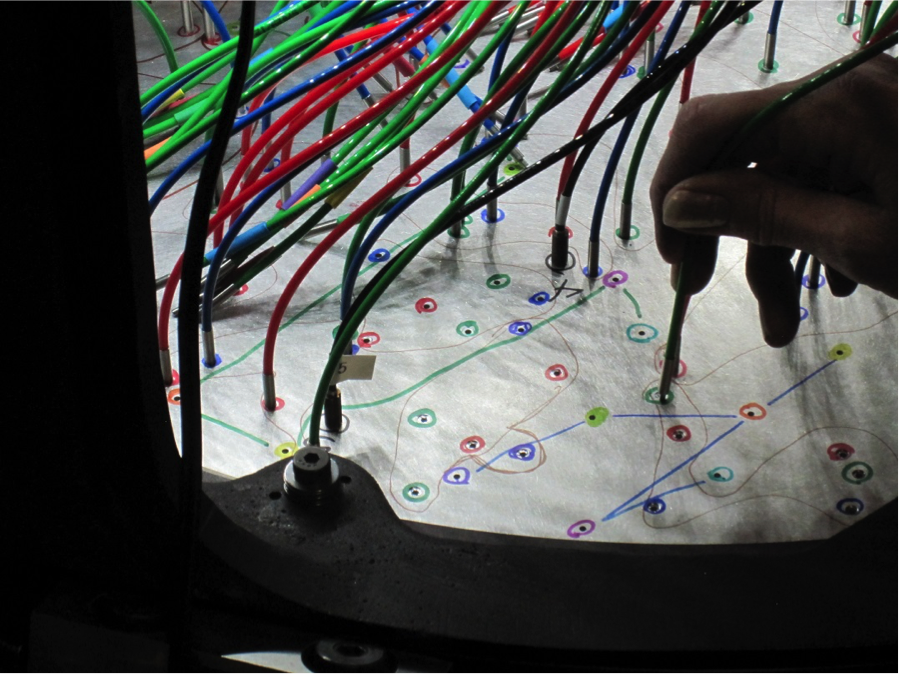}
    \caption{Focal plane plate plugged with optical fibres at pre-drilled holes \cite{SDSS4_Instruments}.}
    \label{fig:plate}
\end{figure}

Historically, MOS observations relied on pre-drilled aluminum plates manually loaded with optical fibers in the telescope focal plane before each observing session (Figure~\ref{fig:plate}) \cite{SDSS_drilling}. Although effective, this approach is inherently time-consuming and inflexible: reconfiguration between targets requires physical intervention, and operational costs scale unfavorably with increasing survey size and target density \cite{Horler2018_HighDensityFP}.

\begin{figure}[H]
    \centering
    \includegraphics[width=0.6\textwidth]{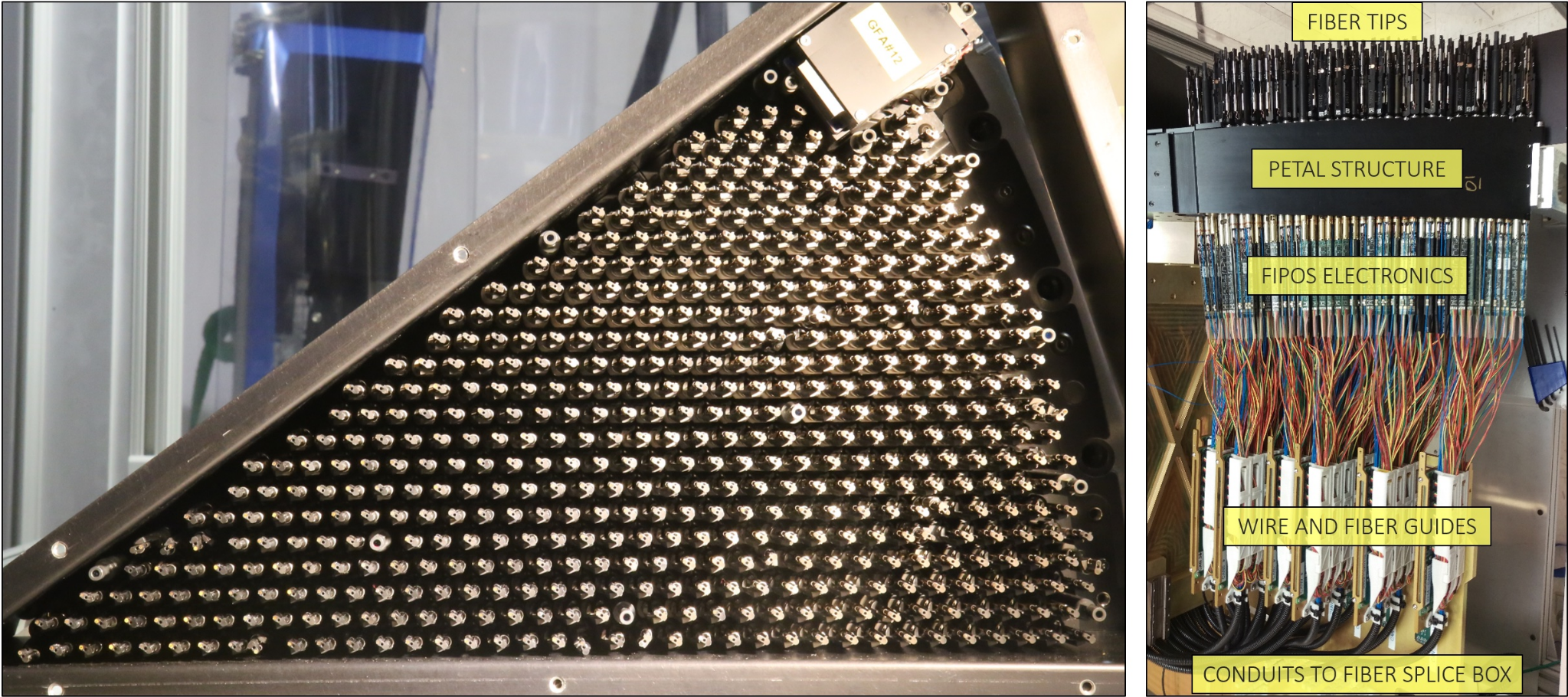}
    \caption{DESI focal plane populated with fibre positioning robots, and the back-end fibre routing and electronics \cite{Silber_2022}.}
    \label{fig:desi}
\end{figure}

To overcome these limitations, modern spectroscopic instruments increasingly employ robotic, adaptive fibre positioners (Figure~\ref{fig:desi}). Each positioner autonomously places an optical fibre within a defined patrol region at high precision, allowing the focal plane to be reconfigured between observations without manual intervention. Complete coverage of a telescope focal plane typically requires on the order of 10,000 to 30,000 such positioners \cite{zhao2025multiplexedsurveytelescopeperspectives}, as deployed, for example, in the Dark Energy Spectroscopic Instrument (DESI) \cite{Silber_2022}. This scale introduces substantial challenges in mechanical design, electronics integration, power distribution, cabling complexity, and large-scale manufacturability.

\subsection{Evolution of control electronics architectures}
\label{sec:evolution}

\begin{figure}[H]
    \centering
    \includegraphics[width=0.8\textwidth]{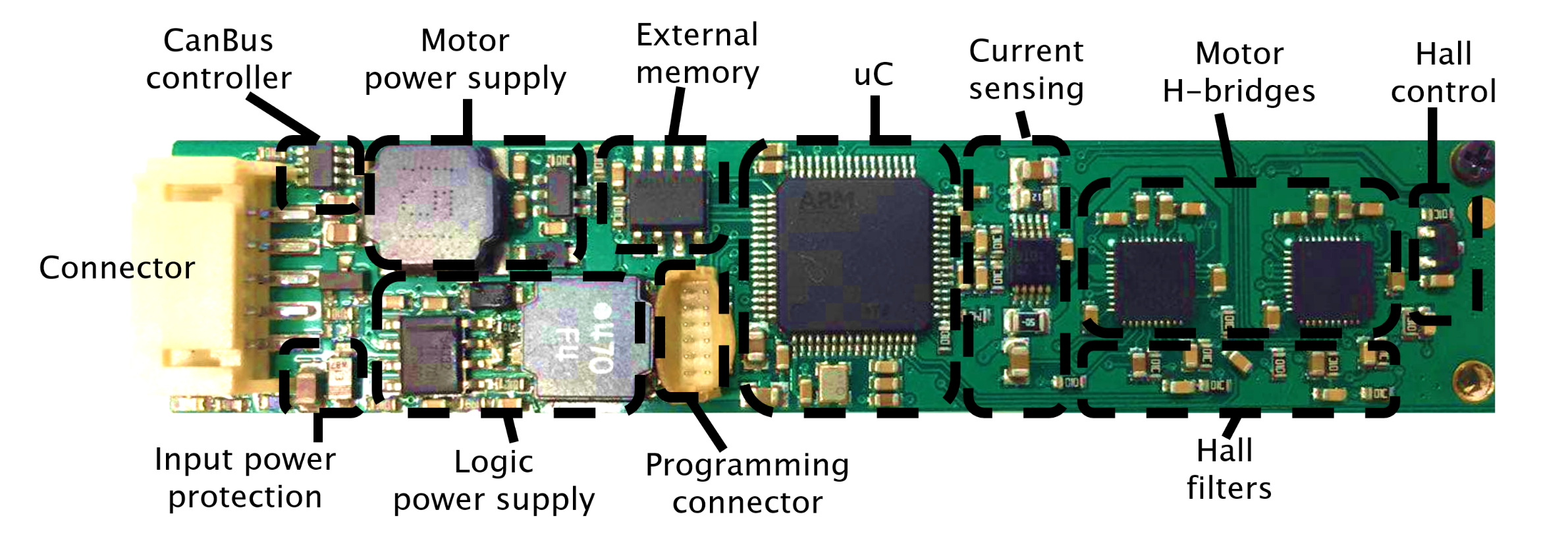}
    \caption{SDSS-V generation electronic card controlling a single positioner \cite{Luzius}.}
    \label{fig:SDSScard}
\end{figure}

The control electronics developed for SDSS-V represent the state of the art against which the present work is benchmarked. In that architecture, each positioner is served by a dedicated printed circuit board (Figure~\ref{fig:SDSScard}) that contains the complete set of functions per-positioner: an STM32F4 micro-controller, a CAN bus controller and transceiver, motor power supply, H-bridge drivers for two BLDC axes, per-phase current detection with Hall filters, external flash memory, input power protection, and a programming connector. This design philosophy — one self-contained electronic node per mechanical actuator, offers clear advantages at small scales: electrical isolation between positioners is implicit, firmware development and debug are straightforward, and board-level failures are localised.

However, the one-board-per-positioner model imposes a set of scaling penalties that become prohibitive as positioner counts grow toward the tens of thousands required by instruments such as MUST and WST. First, the PCB footprint of each board must fit within the mechanical pitch of the positioner array, directly constraining components choice and routing density. Second, the wiring harness, power, CAN, and signal lines routed individually to each board grows linearly with positioner count, imposing a proportional increase in assembly time, connector count, and potential failure sites. Third, each board duplicates fixed-overhead circuitry (power regulation, CAN transceivers, oscillator) that could otherwise be shared across many actuators, leading to unnecessary silicon area, board area, and quiescent power consumption. 

Finally, per-unit manufacturing and test costs remain high regardless of the production volume, since no component is shared between boards. These limitations were directly observed during the operation of DESI-generation instruments and motivated the architectural transition described in Section~\ref{sec:New Modular Approach}. 

The fundamental requirement driving the redesign is the need to achieve a 21-fold reduction in board count while preserving full per-positioner real-time autonomy, a constraint that shapes every aspect of the new architecture.

The Dark Energy Spectroscopic Instrument (DESI) advanced beyond the fully discrete model by grouping ten positioners onto a single petal controller, but the motor drive stage remains essentially one H-bridge set per robot, with dedicated current-sense paths and per-positioner CAN addressing handled by daisy-chained transceivers along each petal. The DESI electronics demonstrated reliable 5,000-positioner operation, yet the architecture still scales linearly in board count and wiring harness mass with positioner number. For instruments targeting 20,000–30,000 fibers, such as MUST and WST, a qualitatively different integration density is required.

\begin{figure}
    \centering
    \includegraphics[width=0.5\linewidth]{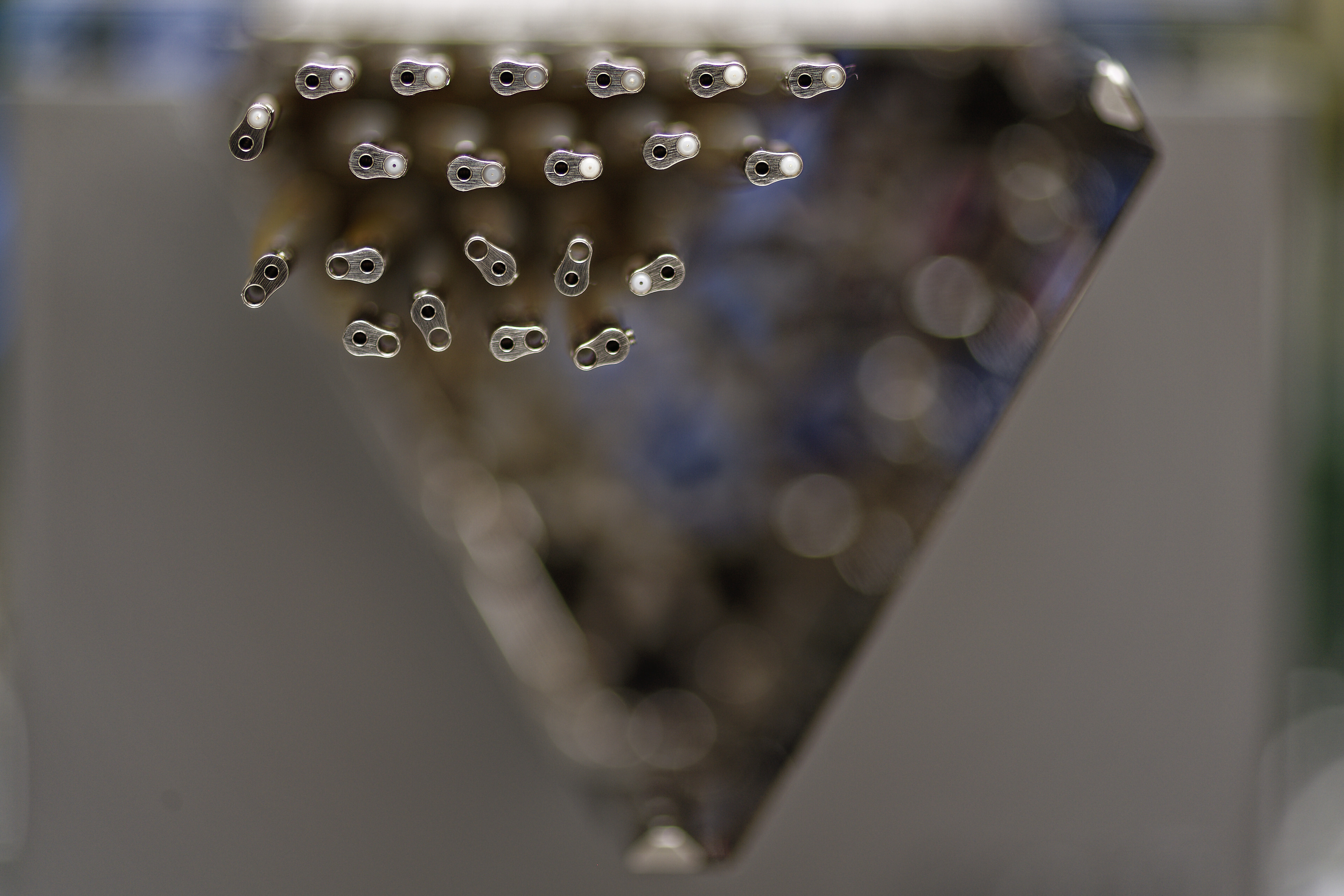}
    \caption{Array of 21 positioners in a 63 triangular module chassis}
    \label{fig:ArrayPos}
\end{figure}

Current-generation architectures pursue significantly higher levels of integration: a single control board managing 21 SCARA-type positioners (42 brushless DC motors), enabling triangular assemblies of 63 units with just three boards. This 21-fold densification improves system-level energy performance while preserving sub-\SI{5}{\micro\metre} positioning accuracy, a requirement for the next generation of massive spectroscopic surveys, including the Multi-fibre Spectroscopic Telescope (MUST) and ESO's Wide-field Spectroscopic Telescope (WST) (Figure~\ref{fig:ArrayPos}).

This work builds on the PhD thesis of Luzius Kronig \cite{Luzius}, developed at Astrobots, which addressed previous generations of spectroscopic instrumentation, including metrology systems, performance metrics, electronics design, and system modeling. The firmware development for the new generation of positioners is documented in the unpublished Master's thesis of Tarik Ibrahimovic \cite{Tarik_thesis}.

\subsection{Objectives and scope of this article}
\label{sec:objectives}

This article presents the development of embedded firmware and a detailed analysis and optimisation of a multi-positioner control board for BLDC-driven SCARA fibre positioning robots. The work translates system-level requirements into concrete electronic hardware and software, delivering a complete platform for motor driving, sensing, communication, firmware updating, and power management.

The principal contributions are as follows:

\begin{itemize}
    \item \begin{sloppypar}\textbf{A 21-positioner shared control board}, that consolidates motor drive, current sensing, CAN communication, and power conversion onto a single PCB, reducing board count by 21× relative to per-positioner architectures while preserving per-MCU real-time independence.\end{sloppypar}
    \item \textbf{Sensorless FOC with current-based hard-stop calibration}, eliminating Hall sensors and encoders through a single-shunt DC-link measurement and a deterministic datum algorithm that achieves  $\leq50 ~\si{\micro\meter}$ repeatability.
    \item \textbf{CAN bootloader for field-scale firmware deployment}, enabling over-the-bus binary updates across ~1,500 boards without physical access.
    \item \textbf{Performance characterization} of positioning accuracy, power consumption, and thermal behavior on a 21-positioner prototype, with identification of open items toward meeting the static power target.
    \item \textbf{Production and testing}:  an integration and test plan to ensure the quality of manufactured assemblies.
\end{itemize}

A key hardware transition with respect to previous generations is the adoption of an STM32G4 micro-controller, replacing the STM32F4-based platform. This newer device provides hardware floating-point support, higher-resolution timers, and on-chip trigonometric accelerators that are well-suited to real-time motor control.

\section{SYSTEM ARCHITECTURE}

\subsection{New Modular Approach}\label{sec:New Modular Approach}

\begin{figure}[H]
    \centering
    \includegraphics[width=1\linewidth]{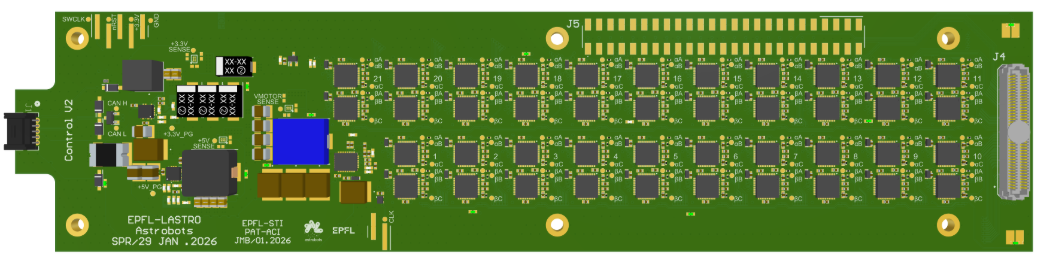}
    \caption{Control PCB top view}
    \label{fig:TopControlV2}
\end{figure}

\begin{figure}[H]
    \centering
    \includegraphics[width=1\linewidth]{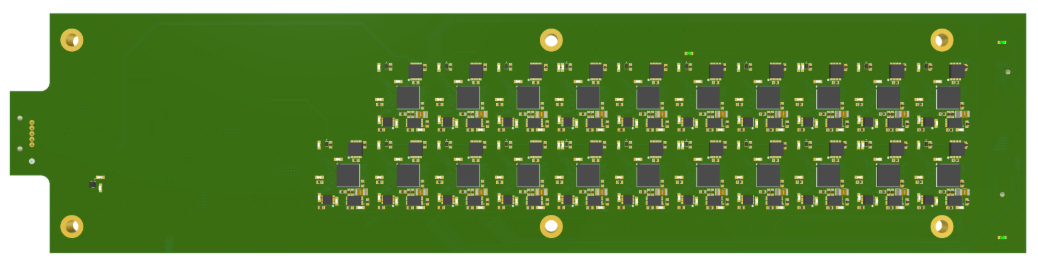}
    \caption{Control PCB bottom view}
    \label{fig:BottomControlV2}
\end{figure}

\begin{figure}[H]
    \centering
    \includegraphics[width=0.5\linewidth]{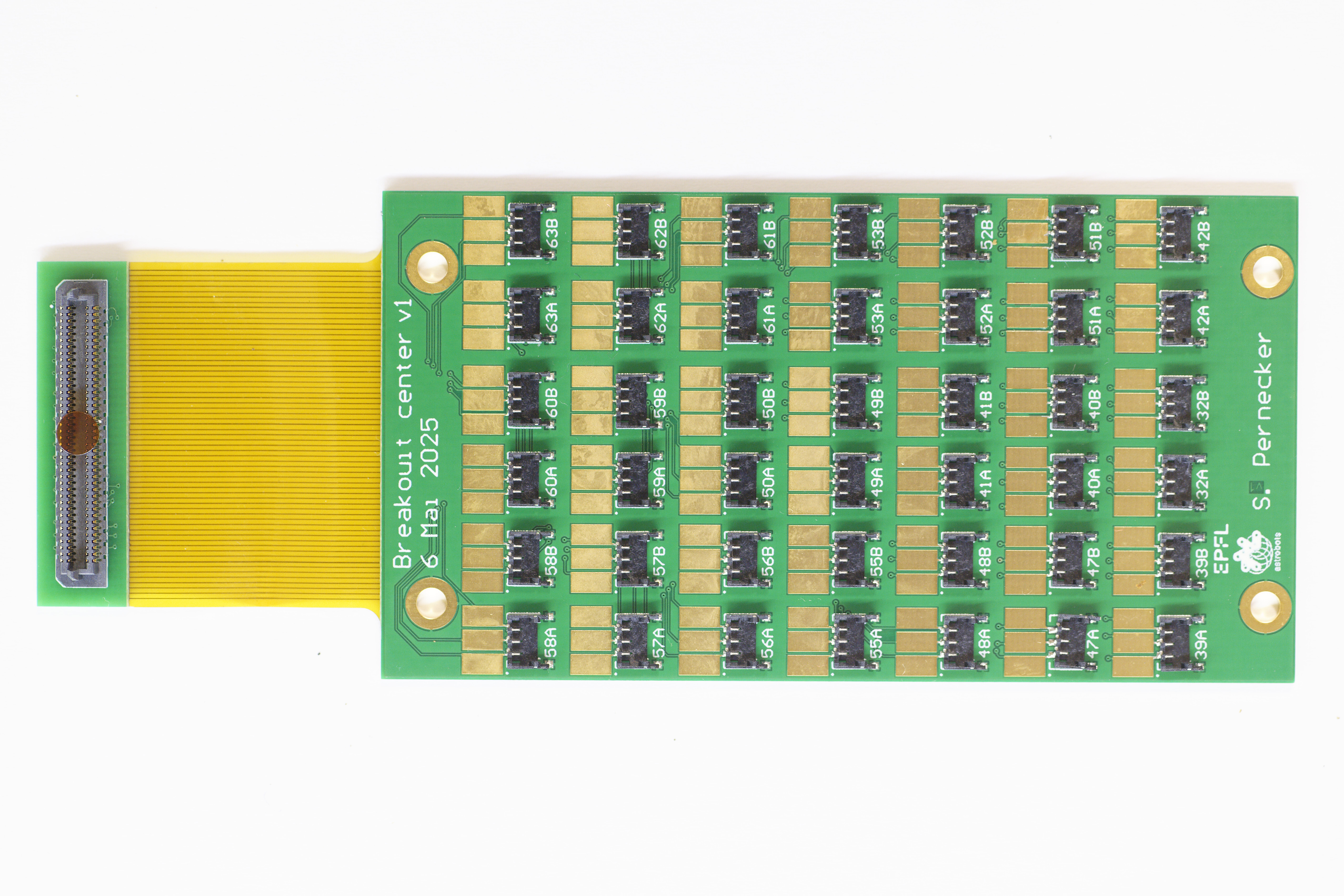}
    \caption{Breakout PCB top view}
    \label{fig:Breakout}
\end{figure}

\begin{figure}[H]
    \centering
    \includegraphics[width=1\linewidth]{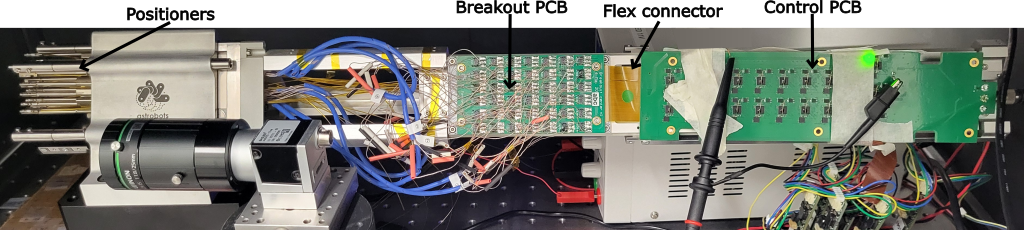}
    \caption{Triangular module with the identification of parts}
    \label{fig:setup}
\end{figure}

To overcome the limitations of the per-positioner PCB architecture, we introduce a modular control strategy organized around a triangular mechanical unit housing 63 positioners (Figure~\ref{fig:setup}). Each module is served by three identical control PCBs (Figure~\ref{fig:TopControlV2} and ~\ref{fig:BottomControlV2} ), with each board driving 21 positioners simultaneously via a Breakout PCB where motor wires are soldered or connected with a small three positions connector (Figure~\ref{fig:Breakout}). This $3 \times 21 = 63$ decomposition defines the fundamental integration unit of the system and directly matches the mechanical tiling geometry of the focal plane.

\begin{figure}[H]
    \centering
    \includegraphics[width=1\linewidth]{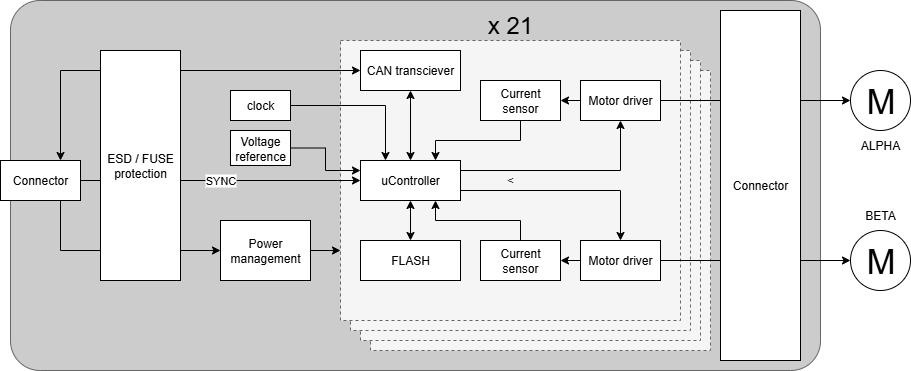}
    \caption{Architecture block diagram}
    \label{fig:ArchitectureBlockDiagram}
\end{figure}

The key architectural decision is the consolidation of motor drive, current sensing, communication, and real-time control resources onto a single shared PCB spanning 21 positioners, rather than replicating a dedicated board at every positioner site. Each positioner retains its own dedicated micro-controller, preserving real-time independence, deterministic CAN responsiveness, and per-positioner fault isolation. Shared resources — power conversion, ESD protection, CAN transceiver, and flash memory, are distributed once per board rather than once per positioner, substantially reducing component count, PCB surface area, and wiring harness complexity (Figure~\ref{fig:ArchitectureBlockDiagram}).

At the system level, three control PCBs are assembled into a single 63-positioner module via a flex connector and breakout PCB interface. This hierarchical structure enables independent board-level testing and replacement before module integration, and supports the scaling requirements of instruments such as MUST ($>$20\,000 positioners) and WST by assembling the required positioner count from a single, validated board design without per-instrument redesign.

\section{CONTROL ELECTRONICS DESIGN}
\subsection{Multi-Positioner Board Architecture}
\begin{figure}[H]
    \centering
    \includegraphics[width=1\linewidth]{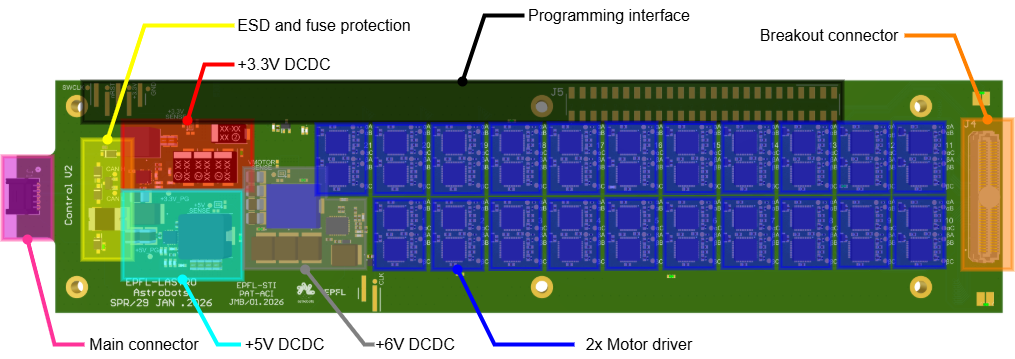}
    \caption{Control PCB top view}
    \label{fig:ControlTop}
\end{figure}

\begin{figure}[H]
    \centering
    \includegraphics[width=1\linewidth]{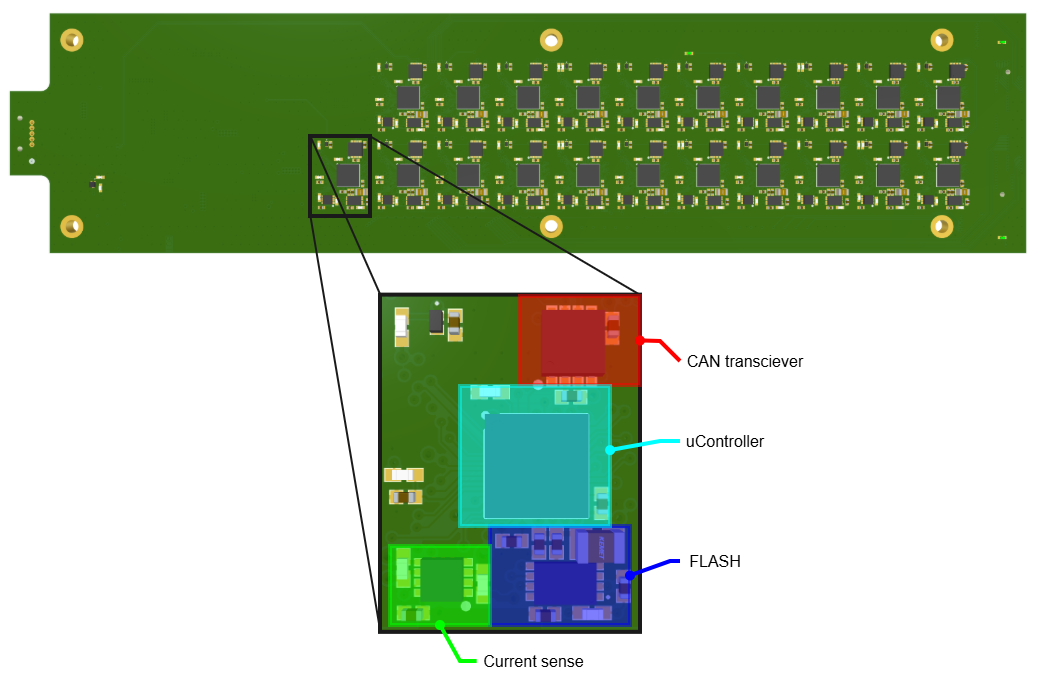}
    \caption{Control PCB bottom view}
    \label{fig:ControlBottom}
\end{figure}

The Control PCBA uses a two-sided components placement. The bottom board carries the half-bridge ICs, DC-DC converters (6V, 5V and 3.3V), and ESD protection. The top board hosts the 21 microcontrollers, an external flash, INA instrumentation amplifiers, and the CAN transceiver. This separation isolates PWM switching noise from sensitive analog and communication paths (Figure~\ref{fig:ControlTop} and Figure~\ref{fig:ControlBottom}).

Each positioner is assigned one dedicated MCU, running its own 2\,kHz control loop independently of the others. All 21 MCUs share a single CAN transceiver and are addressed individually by node ID. CAN commands are decoded asynchronously, with acknowledgement frames sent immediately — responsiveness across all 21 nodes has been validated. Motor drive signals and current sense returns are routed to the positioner array through a flex connector and breakout PCBA, following the signal chain:
$\text{MCU} \rightarrow \text{half-bridge} \rightarrow \text{PMSM} 
\rightarrow \text{shunt} \rightarrow \text{INA} \rightarrow \text{ADC} 
\rightarrow \text{MCU}$.

\subsection{Bootloader and Firmware Application}\label{sec:Bootloader and Firmware Application}

Each microcontroller requires dedicated embedded software to operate the 
hardware platform. This software is structured as two distinct and independent 
programs residing in separate regions of the MCU's internal flash memory: 
the \emph{bootloader} and the \emph{application firmware}.

\subsubsection*{General design methodology}
Both programs share a common set of design principles aimed at reliability, maintainability, and energy efficiency. Execution is entirely interrupt-driven: the MCU issues a Wait-For-Interrupt (\texttt{WFI}) instruction whenever no active task is pending, suspending the processor clock and reducing dynamic power consumption during the long idle intervals between motor moves. Inter-process communication is handled through shared status flags polled at each interrupt entry, avoiding the overhead of a real-time operating system at the current scale of one MCU per positioner. Persistent state, including motor configuration parameters and the last known joint positions, is stored in the MCU's internal flash, ensuring that the system can recover gracefully after a power cycle without requiring a full re-homing sequence. Hardware abstraction is enforced through opaque struct encapsulation for the motor and CAN objects, which isolates the application logic from low-level peripheral details and simplifies future porting to 
revised hardware revisions.

\subsubsection*{Bootloader}
The bootloader runs first on every power-up and serves two purposes: it validates the integrity of the application image before handing over execution, and it provides a field-update path over the CAN bus that requires no physical access to the PCB. On reset, the bootloader reads a boot-status flag from a reserved flash location. If the flag is valid, it waits  for a CAN boot request from the host;  it jumps immediately to the application. If a firmware update command is received over CAN, the bootloader accepts the new binary frame-by-frame, writes it to the application flash partition, and verifies the full image with a CRC-32 checksum before marking the boot-status flag as valid and rebooting. If the flag is found invalid  at power-up — for example on a freshly programmed board or after a corrupted update — the bootloader enters update mode unconditionally, preventing the MCU from ever jumping to a corrupted image. This mechanism is essential at instrument scale: with approximately 1,500 boards deployed across the focal plane, the ability to push firmware updates over the existing CAN infrastructure without module dismounting is a critical operational requirement.

\subsubsection*{Application firmware}
\begin{figure}[H]
    \centering
    \includegraphics[width=1\linewidth]{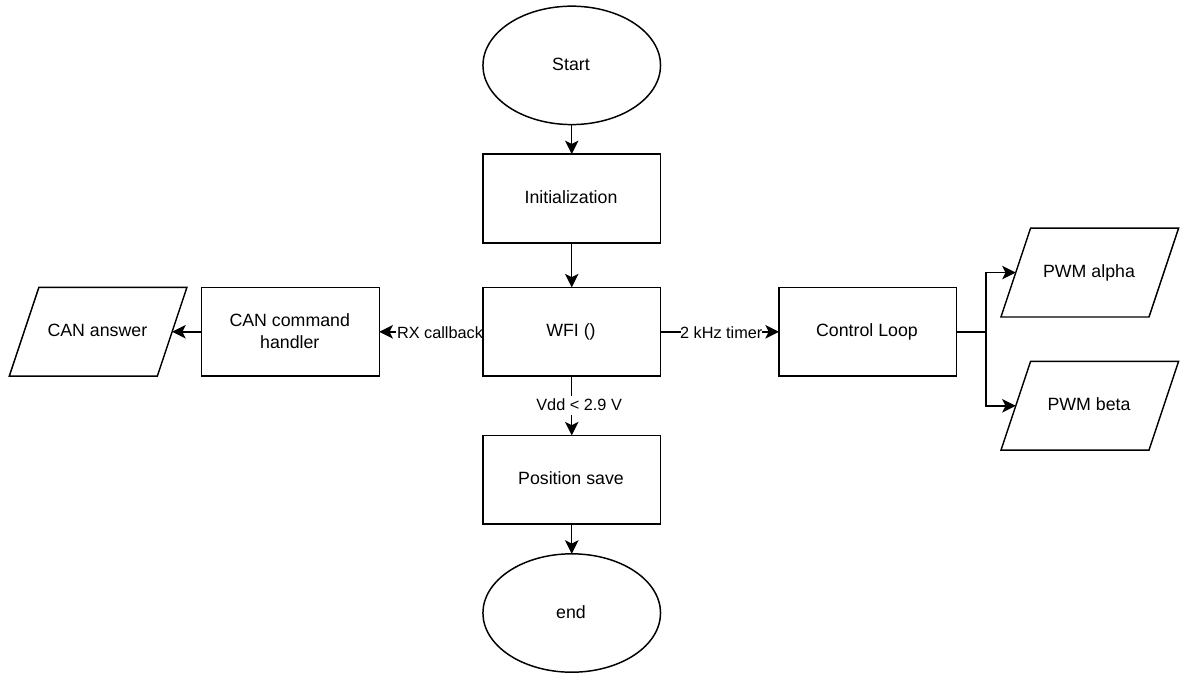}
    \caption{Overall firmware architecture. The MCU sleeps in 
    \texttt{WFI} between events and is woken by three sources: 
    the 2\,kHz control-loop timer, an incoming CAN frame, 
    or a supply undervoltage condition triggering an emergency 
    position save.}
    \label{fig:firmwareArch}
\end{figure}

After the bootloader transfers control, the application initialises all peripherals and enters the main event loop, whose architecture is illustrated in Figure~\ref{fig:firmwareArch}. The MCU spends the majority of its time in \texttt{WFI} sleep, and is woken by exactly two asynchronous sources.

The first is a \textbf{2\,kHz timer interrupt}, which drives the real-time control loop. On each tick the firmware performs, in order: trajectory interpolation to advance the $\alpha$ and $\beta$ target positions along the pre-computed move profile; acquisition of the DC-link current $I_m$ from the ADC for the correlated double-sampling filter and the hard-stop detection algorithm; evaluation of the collision and datum-detection conditions; and finally output of the updated PWM duty cycles to the $\alpha$- and $\beta$-axis half-bridges. If the commanded displacement is complete, the motor voltages are held constant at their last value rather than being zeroed, maintaining rotor lock against external disturbances.

The second interrupt source is an asynchronous \textbf{CAN receive callback}, triggered whenever a frame addressed to this MCU arrives on the bus. The callback decodes the command identifier, updates the relevant control flags or parameters visible to the timer loop (target position, move speed, operational mode, etc.), and immediately enqueues a CAN answer frame to acknowledge the command. Because the callback is asynchronous with respect to the 2\,kHz loop, the bus remains responsive at all times, including during active motor moves.

A third, non-periodic exit from \texttt{WFI} is triggered by a \textbf{supply voltage monitor}: if $V_\mathrm{dd}$ falls below 2.9\,V, indicating an imminent power loss, the firmware performs an emergency position save, writing the current joint angles to flash before the supply collapses. This ensures that datum calibration state is preserved across unplanned power interruptions.

\subsection{SCARA Robot Control}\label{sec:Scara Robot Control}
\begin{figure}[H]
    \centering
    \includegraphics[width=0.6\textwidth]{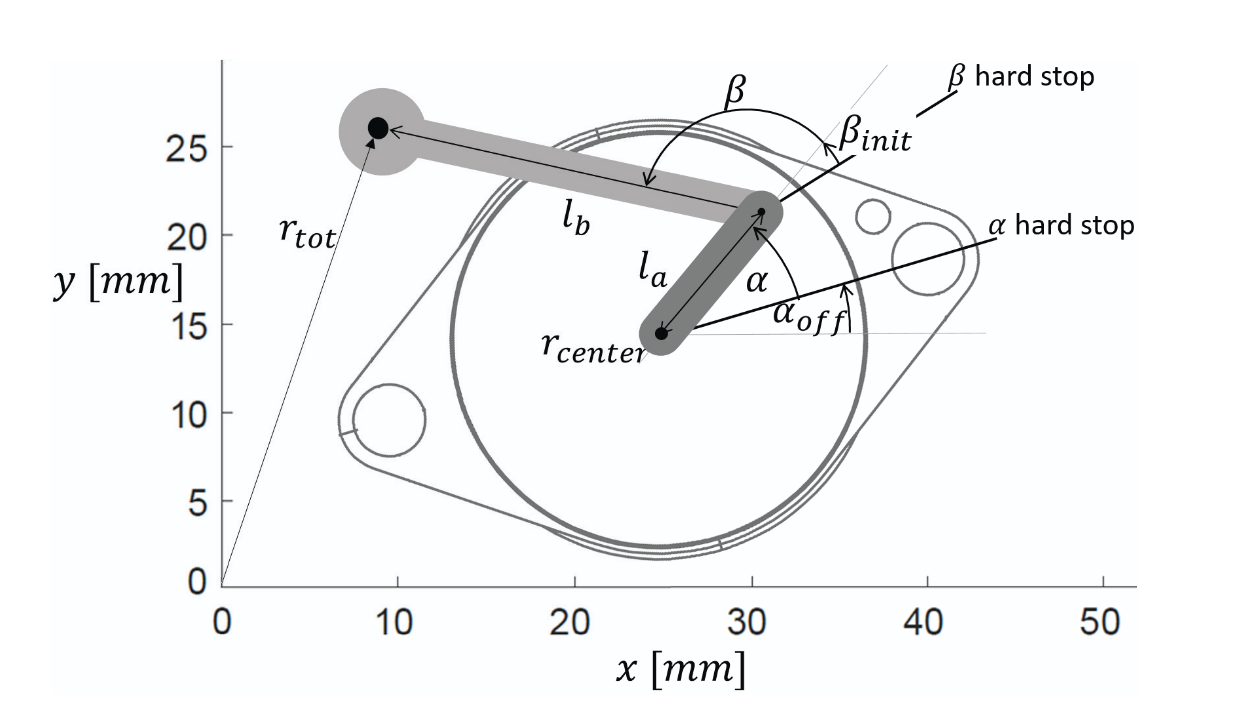}
    \caption{ XY representation of a SCARA positioner with different length for $\alpha$ and $\beta$ arms \cite{Luzius}}
    \label{fig:pos_drawing}
\end{figure}

Each fiber positioner follows a planar theta-phi ($\alpha$-$\beta$) SCARA architecture as represented on Figure~\ref{fig:pos_drawing} , in which two co-axial rotational joints---each with arm lengths $l_\alpha = l_\beta = 1.8$~mm---drive the fiber tip across a circular patrol disk of 7.2~mm diameter. 

Both axes are actuated by 4~mm ironless three-phase 
PMSMs with a 280:1 or 337:1 gear ratio (depending on the motor manufacturer Maxon or Orbray), and each motor is controlled independently by a dedicated MCU on the shared control board.

\subsubsection*{Motion Control Algorithms}
Motor commutation is performed using open-loop Field-Oriented Control (FOC), deliberately forgoing closed-loop shaft encoders. This choice is justified by the high gear ratio: any open-loop angular error at the motor output is attenuated by a factor of the gear ratio at the positioner joint, reducing the residual end-effector contribution to a level compatible with the 5~\si{\micro\meter} RMS positioning budget. 

\begin{figure}[H]
    \centering
    \includegraphics[width=\linewidth]{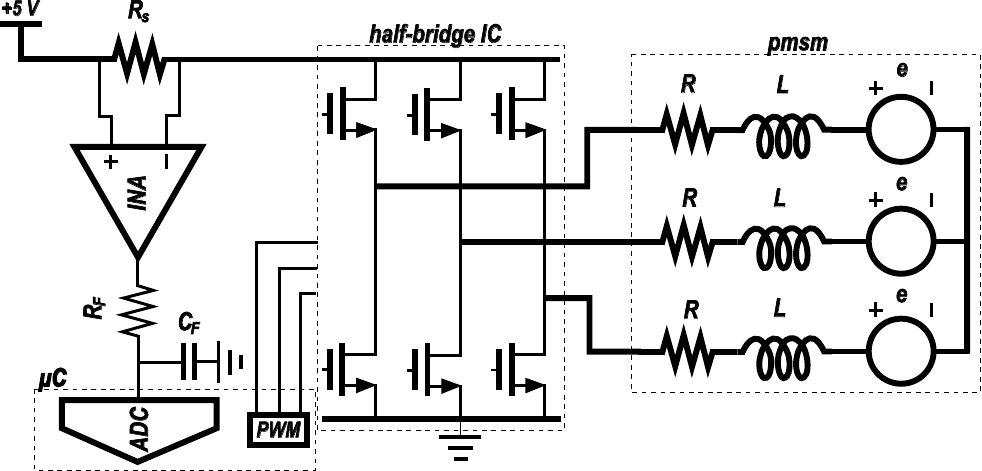}
    \caption{Motor driving and current sensing system schematic}
    \label{fig:modelschem}
\end{figure}

The firmware runs an interrupt-driven control loop at 2~kHz. On each tick, the loop checks whether the commanded displacement has been completed; if not, it performs time-based trajectory interpolation to update the $\alpha$  and $\beta$ target angles, fetches DC-link current measurements from the ADC, runs collision and datum detection, and drives the PWM outputs for both motor axes. Before any positioning can take place, the absolute angular reference of each motor must be established. Figure~\ref{fig:modelschem} represents the hardware responsible to drive the H-Bridge and the ADC measurement.

\begin{figure}[H]
    \centering
    \includegraphics[width=0.5\textwidth]{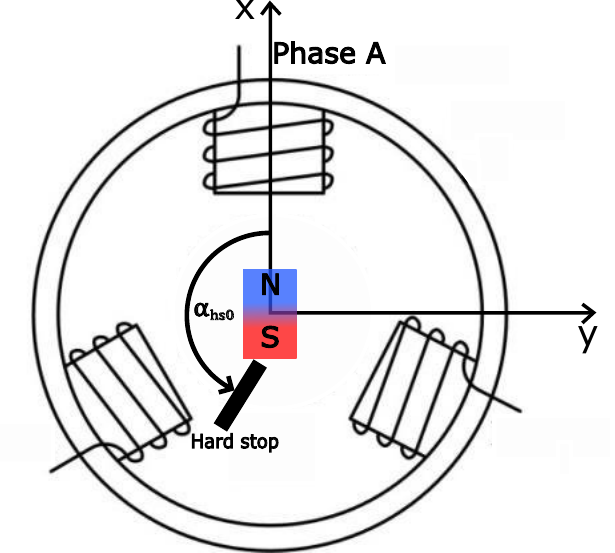}
    \caption{PMSM motor internals with reference to the external hardstop position}
    \label{fig:motor_and_hardstop}
\end{figure}

Because the system operates without encoders, the rotor orientation relative to the mechanical hard-stop, parameterised as $\alpha_{\mathrm{hs0}}$, is unknown after power-up. Without this calibration, end-effector positioning errors can reach 5–13~\si{\micro\meter} across the workspace. Figure~\ref{fig:motor_and_hardstop} illustrates this concept, where $\alpha_{hs0}$ denotes the rotor-to–hard-stop angle when the motor is aligned with the reference phase A.

\begin{figure}[H]
    \centering
    \includegraphics[width=0.75\linewidth]{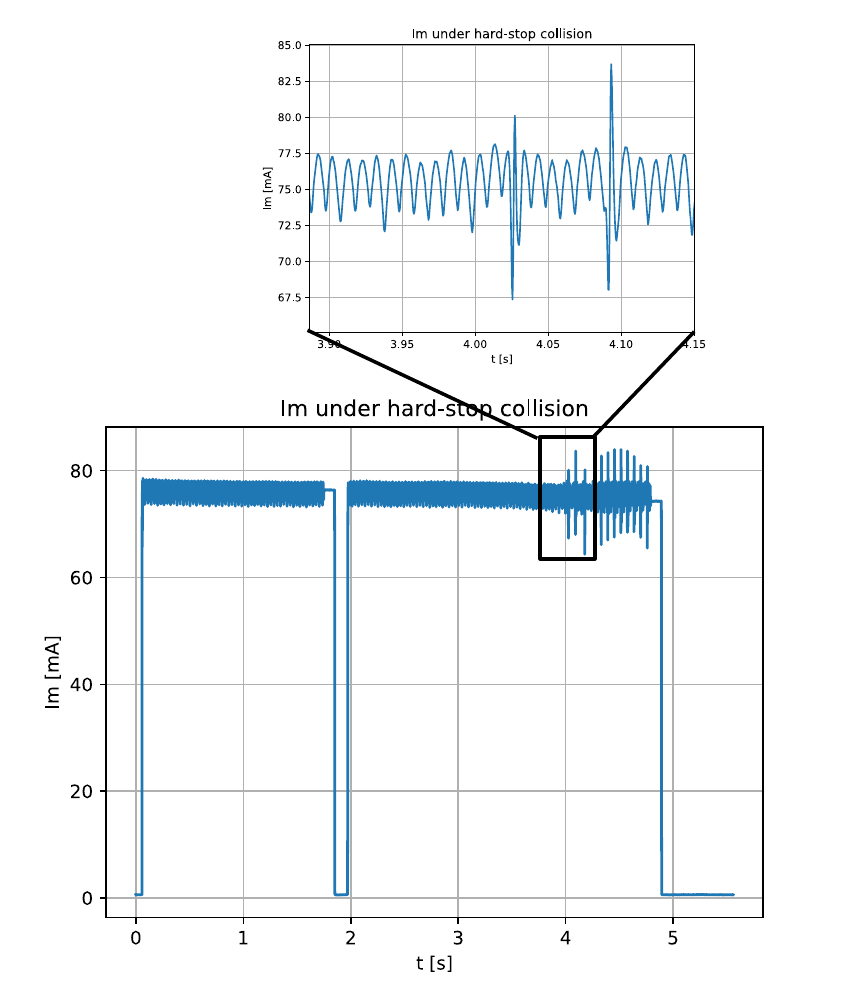}
    \caption{$I_m$ under collision conditions}
    \label{fig:collisionCurrent}
\end{figure}

The datum procedure commands both axes to $-360^{\circ}$ until the rotor stalls against the hard-stop. The resulting characteristic peak in the DC-link current $I_m$ as shown on Figure~\ref{fig:collisionCurrent}. This peak current is detected in real time  ; the stator electrical angle at that instant is recorded and used to compute $\alpha_{\mathrm{hs0}}$, which is then stored in flash memory. 

This sensor-free calibration technique eliminates the need for any additional position-sensing hardware and achieves a datum repeatability within the 50~\si{\micro\meter} specification.

\subsubsection*{Camera Positioning Feedback}
Open-loop FOC alone is not sufficient to meet the 5~\si{\micro\meter} end-to-end accuracy requirement at the fiber tip, owing to residual mechanical imperfections such as gear backlash, arm-length tolerances, and thermal drift. A Fiber View Camera (FVC) therefore provides the outer closed-loop correction. After all positioners in a module have executed their open-loop moves concurrently, the fiber tips are back-illuminated and imaged by the FVC. Centroid positions are extracted by 2D Gaussian fitting and compared to the target sky coordinates projected onto the focal plane. 

\begin{figure}[H]
  \makebox[\textwidth][c]{%
    \begin{minipage}{0.7\textwidth}
      \centering
      \captionsetup[subfigure]{justification=centering}
      \begin{subfigure}[b]{0.45\textwidth}
        \centering
        \includegraphics[width=\linewidth]{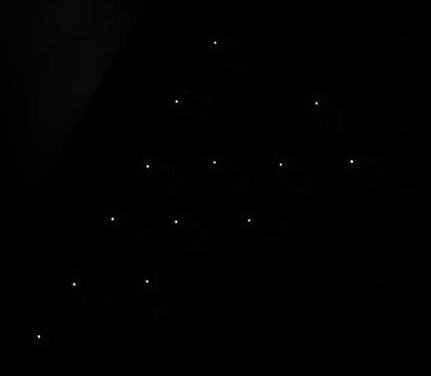}
        \caption{}
        \label{fig:protomps}
      \end{subfigure}
      \hspace{.3cm}
      \begin{subfigure}[b]{0.42\textwidth}
        \centering
        \includegraphics[width=\linewidth]{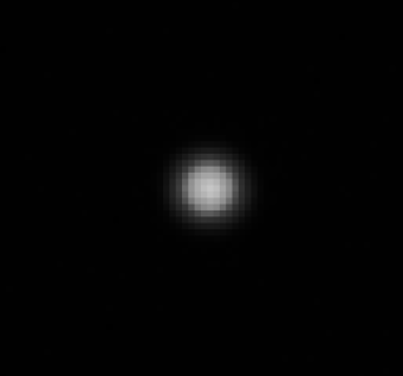}
        \caption{}
        \label{fig:blob_zoomedin}
      \end{subfigure}
    \end{minipage}
  }
  \vspace{.1cm}
  \caption{Light blobs captured by the camera; (a) image captured by the camera showing an array of backlit fibers; (b) Zoomed-in version of one of these blobs}
  \label{fig:multi-blob}
\end{figure}

Correction offsets are computed and sent to each positioner as an incremental angular command. This metrology-correction cycle typically converges in 1 to 3 iterations for the target, yielding a total reconfiguration time well below 40~seconds. In the current EPFL design, the on-board firmware can store up to 50 pre-computed, collision-free trajectories per positioner, enabling the host software to pipeline target assignment and correction without additional communication overhead.

\subsubsection*{Collision Avoidance in High-Density Configuration}
In a high-density focal plane, adjacent positioner patrol disks overlap by design to maximize sky coverage. This overlap creates the possibility of mechanical collisions between neighboring arms during reconfiguration. Two complementary strategies are employed.
 
\paragraph{Trajectory-level avoidance}
Before any reconfiguration, a host-side path planner computes collision-free trajectories for all positioners in a module using a decentralized priority-based algorithm. Each positioner receives its trajectory as a sequence of angular waypoints stored in on-board flash (up to 50~segments per axis). Because all 21~MCUs on a board share a hardware synchronization line (active-low pulse), their trajectory playback starts simultaneously, ensuring that the pre-computed collision-free timing assumptions hold during execution.
 
\paragraph{Current-based runtime detection.}
As a second line of defense, the firmware monitors the DC-link current~$I_m$ during every \SI{2}{\kilo\hertz} control tick. A sustained current exceedance above a configurable threshold (distinct from the transient peak used for hard-stop datum) triggers immediate motor braking on the affected axis and raises a fault flag visible to the host over CAN. This mechanism cannot prevent contact entirely---the detection latency is one control period (\SI{500}{\micro\second}) plus mechanical inertia---but it limits the collision energy and protects the positioner gearing. The threshold is set above the maximum expected load current during normal operation to avoid false positives, and below the stall current to ensure detection before mechanical damage. Experimental characterization of the detection latency and residual collision force is ongoing.

\section{PERFORMANCE ANALYSIS}
\subsection{Positioning Accuracy}
The system targets a fiber positioning repeatability of $\leq 5\,\mu\mathrm{m}$ RMS at the focal plane, consistent with the astrometric precision budget of next-generation instruments. This requirement is demanding in the context of open-loop motor control: with arm lengths $l_\alpha = l_\beta = \SI{1.8}{\milli\metre}$ and a gear ratio of 280:1 or 337:1, the allowable angular error per axis at the motor shaft is on the order of a few tenths of a degree.

\subsubsection*{Open-loop performance and the role of gear reduction}
As described in Section~\ref{sec:Scara Robot Control}, each PMSM is driven in open-loop FOC without shaft encoder or Hall sensor, relying on the high gear ratio to attenuate motor-level angular errors at the end-effector. Simulations with a validated LTspice model confirm that a nominal load angle $\delta_L = 10^\circ$ produces end-effector displacements well within the $5\,\mu$m budget under nominal operating conditions.

\subsubsection*{Impact of initial motor orientation and hard-stop calibration}
Without the hard-stop datum calibration described in Section~\ref{sec:Scara Robot Control}, the motor orientation $\alpha_{hs0}$ is unknown at power-up. Monte Carlo analysis over the full $\pm 40^\circ$ uncertainty range of $\alpha_{hs0}$ demonstrates worst-case end-effector errors of 5--13\,$\mu$m across the two-dimensional workspace, sufficient to violate the positioning requirement at the edges of the distribution. Following calibration, positioning errors are reduced to within the $5\,\mu$m requirement in simulation, a result currently being validated experimentally.

\subsubsection*{Datum repeatability}
Datum (hard-stop) repeatability---the ability to return to the reference position consistently across successive homing cycles---is a prerequisite for closed-loop correction using the Fiber View Camera (Section~\ref{sec:Scara Robot Control}). The $\leq 50\,\mu$m datum repeatability requirement has been met experimentally across the 21-positioner prototype board. At the system level, 1--3~FVC correction iterations converge the final fiber position to within the $5\,\mu$m astrometric budget in under 40\,s of reconfiguration time.

\subsubsection*{Residual error sources}
The dominant contributors to the residual positioning error budget are: (i) mechanical backlash in the positioner gear train, which introduces a systematic offset between approach directions and must be mitigated by unidirectional motion profiling; (ii) thermal gradients across the focal plate, which are kept below a $10^\circ$C rise to limit thermally induced defocus to $\leq 5\,\mu$m; and (iii) position estimator noise, bounded experimentally to $\pm 40^\circ$ electrical, equivalent to a sub-degree mechanical error after gear reduction. Full hardware validation of end-to-end positioning repeatability across the 21-positioner array is ongoing.

\subsection{Energy Efficiency and Thermal Management}

Power dissipation is a first-order constraint for dense robotic focal planes: heat generated by the control electronics is conducted through the PCB stack and module frame directly into the focal plate, where any temperature rise maps to a thermomechanical positioning error. For the MUST focal plane, a maximum $10^\circ\mathrm{C}$ temperature rise is imposed to keep thermally induced defocus within the $5\,\mu\mathrm{m}$ positioning error budget. This sets two hard requirements per 21-positioner board: $\leq 0.5\,\mathrm{W}$ in the static (idle) regime and $\leq 50\,\mathrm{W}$ during active reconfiguration.

The board is powered from a single \SI{12}{\volt} external supply routed through a reverse-polarity protection MOSFET and an input EMI filter. Three onboard DC--DC converters derive the internal rails: \SI{6}{\volt} for the motor half-bridges, \SI{5}{\volt} for the CAN transceiver and external flash, and \SI{3.3}{\volt} for the MCU cluster and instrumentation amplifiers. In the static regime---the dominant state during $\sim$\SI{15}{\minute} science exposures---all 21~MCUs execute WFI sleep between \SI{2}{\kilo\hertz} control ticks with motor outputs at zero duty cycle. In the dynamic regime, all 42~motors may be commutated simultaneously for up to \SI{40}{\second}, with dissipation driven by half-bridge switching losses and motor copper losses. The predecessor SDSS-V architecture duplicated power regulation, CAN transceivers, and memory at every positioner; at counts exceeding 20\,000 this becomes untenable. The new architecture shares these resources across 21~positioners, reducing per-positioner passive overhead by a factor of~21 and projecting total focal-plane idle power to approximately \SI{2.3}{\kilo\watt} across $\sim$1\,500~boards---compatible with passive conductive cooling without active refrigeration.

The current prototype draws $\sim$\SI{1.5}{\watt} at idle, exceeding the static target by a factor of three. The dominant contributors are the quiescent current of the DRV8316 half-bridge ICs and the MCU active-mode baseline between WFI entries; mitigation paths under evaluation include lower-quiescent half-bridge substitutes, supply-rail gating via a load switch during exposures, and deeper MCU sleep modes (Stop~1). The static specification is under review with the MUST and WST instrument teams. In the dynamic regime, peak power at full current across all 21~positioners is $\sim$\SI{30}{\watt}, within the \SI{50}{\watt} requirement; typical operational power is significantly lower, as motors are not simultaneously at peak current across all axes throughout a move.

\begin{figure}[H]
    \centering
    \includegraphics[width=0.5\linewidth]{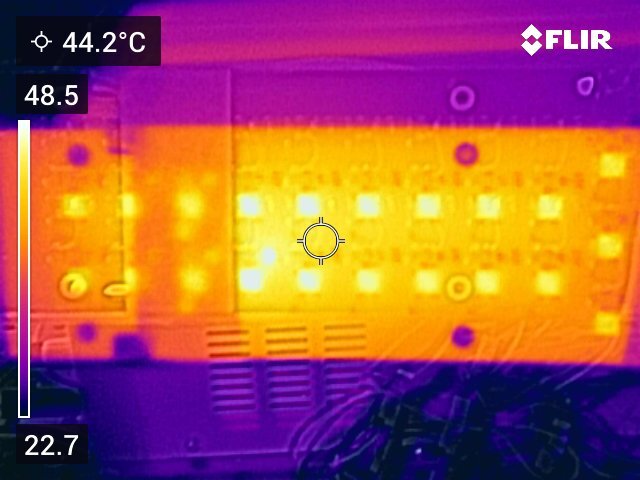}
    \caption{Thermal image of the half-bridge motor drivers during active positioner commutation.}
    \label{fig:Thermal}
\end{figure}

During dynamic operation the primary heat sources are the half-bridge ICs on the lower board; locating them opposite the logic components facilitates heat extraction toward the module chassis (Figure~\ref{fig:Thermal}). Heat is removed conductively through the rigid-flex stack into the aluminium frame and from there into the \SI{1500}{\milli\metre}-diameter focal plate. No active cooling is foreseen at the module level. The firmware includes multi-point thermal monitoring to flag boards approaching the thermal envelope and to interleave move sequences to redistribute dissipation in time. Full characterisation of the static and dynamic thermal impedance from PCB to focal plate is planned as part of the module-level integration campaign.

\subsection{Scalability}
The modular 3×21 architecture is designed to tile without per-instrument redesign. The fundamental scaling unit is the 63-positioner triangular module; an instrument focal plane is populated by replicating this module across the available area. For MUST (~20,000 positioners), approximately 318 modules comprising 952 control boards are required. For WST (~32,000 positioners), the count rises to roughly 476 modules and 1,428 boards. Because each board is functionally identical — differing only in CAN node-ID assignment configured at first boot — manufacturing and qualification follow a single BOM and test procedure regardless of instrument scale.
The CAN bus is the system-level communication backbone. Each 21-positioner board exposes a single CAN transceiver to the instrument bus; at the module level, three transceivers share a common bus segment. The host controller addresses individual MCUs by a hierarchical node-ID scheme: module number, board index (0–2), and positioner index (0–20). Bus load analysis at 1 Mbit/s indicates that a single CAN segment can support up to 12 modules (36 boards, 756 positioners) with sufficient margin for acknowledgement traffic, firmware update frames, and telemetry. Beyond this, additional CAN segments are added with dedicated host interfaces, a standard scaling pattern in automotive and industrial fieldbuses.
The bootloader-based firmware update path (section \ref{sec:Bootloader and Firmware Application}) is critical to scalability: deploying a firmware revision across 1,500 boards takes approximately 45 minutes over CAN at the current frame rate, with automatic CRC-32 verification and fallback to update mode on corruption. No physical access to any board is required.

\subsection{Reliability}
Board-level fault isolation is preserved by the one-MCU-per-positioner mapping. A failed MCU disables a single positioner without affecting its 20 neighbors on the same board or the remaining 42 positioners in the module. If a shared resource fails — the CAN transceiver, power converter, or flash memory — all 21 positioners on that board are lost, but the remaining two boards in the module continue to operate independently. Board replacement is performed at the module level: the affected control PCB is disconnected from the flex interface and swapped without disturbing the positioner mechanics or the other two boards.
The firmware contributes to reliability through three mechanisms: 
(i) emergency position save on supply undervoltage (section \ref{sec:Bootloader and Firmware Application}), preserving datum calibration across unplanned power loss; 
(ii) CRC-protected firmware images preventing execution of corrupted code;
(iii) per-positioner watchdog timers that reset the MCU if the 2 kHz control loop stalls, with automatic re-entry into a safe braked state on recovery.
Mean-time-between-failure (MTBF) estimates for the control electronics are being developed based on component-level reliability data (MIL-HDBK-217) and will be validated through accelerated life testing during the module qualification campaign.

\section{CHALLENGES AND SOLUTIONS}

\subsection{Electromagnetic interference in densified configurations}
Consolidating 42 PWM-driven motor channels onto a single 80\,mm $\times$ 120\,mm PCB concentrates switching noise in a confined area. The primary coupling path is conducted EMI from the half-bridge switching transients into the analog current-sense return and the CAN communication lines. Three mitigation measures are implemented. First, the PCB stackup places a continuous ground plane between the power (bottom) and signal (top) layers, providing a low-impedance return path that limits common-mode loop area. Second, the INA instrumentation amplifiers used for DC-link current sensing provide high common-mode rejection ($>$80\,dB at the PWM frequency), attenuating differential noise induced by adjacent channel switching. Third, CAN bus termination resistors are placed at both ends of each segment, and the differential signaling of the CAN physical layer provides inherent immunity to the common-mode noise levels measured on the prototype ($<$200\,mV peak).

Radiated EMI has not yet been formally characterized. Pre-compliance measurements using a near-field probe indicate that emissions are concentrated at the PWM fundamental (20\,kHz) and its harmonics, with levels that decay rapidly beyond 30\,mm from the board edge. Full EMC characterization is planned during module-level integration.

\subsection{Signal integrity at high density}
The 21 MCUs share a single SPI bus to the external flash and a single CAN bus to the host. SPI clock integrity at 8\,MHz over the 120\,mm bus length has been verified by eye-diagram measurement, with adequate noise margin at all tap points. The CAN bus operates at 1\,Mbit/s; bit-timing analysis confirms that the propagation delay across three daisy-chained boards within a module (total stub length $<$400\,mm) is within the CAN 2.0B specification for this bit rate, with a sample point at 75\% providing robust arbitration.

\subsection{Manufacturing and testing at scale}
Each control PCB carries over 600 components on two sides, requiring automated pick-and-place, reflow, and automated optical inspection (AOI). The test plan is structured in three stages: (i)~bare-board electrical test (flying-probe) to verify all nets before population; (ii)~board-level functional test after assembly, exercising each MCU's CAN communication, ADC acquisition, and PWM output into a resistive dummy load; and (iii)~module-level integration test, in which all 63 positioners are driven to known positions and verified against the Fiber View Camera. Stage~(ii) uses a custom bed-of-nails fixture that contacts all 21 programming headers simultaneously, enabling parallel bootloader flashing and functional verification in under five minutes per board.

\section{INTEGRATION PERSPECTIVE}

\subsection{Compatibility with MUST and WST}
The control electronics presented in this work are designed to satisfy the
requirements of two next-generation spectroscopic instruments: the
Multiplexed Survey Telescope (MUST, $>$20\,000 positioners) and ESO's
Wide-field Spectroscopic Telescope (WST, $\sim$32\,000 positioners).
Both instruments impose similar constraints on the electronics: sub-\SI{5}{\micro\metre}
fiber positioning accuracy, a static power envelope compatible with passive
conductive cooling through the focal plate, and full focal-plane
reconfiguration in under \SI{60}{\second}. The modular $3\times21$ architecture
maps directly onto these requirements: the triangular tiling geometry matches
the planned focal-plane layouts, the CAN-based communication backbone
scales to the required positioner counts through bus segmentation
(Section~\ref{sec:Bootloader and Firmware Application}), and the
single-BOM board design avoids per-instrument hardware redesign.
Instrument-specific adaptations are limited to CAN node-ID mapping,
motor parameter tuning for the selected gear ratio, and thermal interface
dimensioning at the module-to-focal-plate boundary.

\subsection{Comparison with international competitors}
Table~\ref{tab:comparison} summarises the key architectural parameters of the
present design alongside those of recent and planned fiber positioner
instruments.

\begin{table}[H]
\centering
\caption{Comparison of control electronics architectures across fiber
positioner instruments.}
\label{tab:comparison}
\begin{tabular}{l c c c c c}
\hline
\textbf{Parameter} & \textbf{SDSS-V} & \textbf{DESI} & \textbf{4MOST} & \textbf{PFS} & \textbf{This work} \\
\hline
Positioners per board     & 1      & $\sim$10  & 1       & 1      & 21 \\
Total positioners         & 500    & 5\,000    & 2\,436  & 2\,394 & 20\,000+ \\
MCU per positioner        & 1      & shared    & 1       & shared & 1 \\
Shaft encoder             & No     & No        & No      & No     & No \\
Current sensing           & Per-phase & Per-phase & ---   & ---    & Single-shunt \\
CAN transceivers / board  & 1      & 1         & 1       & ---    & 1 \\
Field firmware update     & No     & Partial   & ---     & ---    & CAN bootloader \\
\hline
\end{tabular}
\end{table}

The principal differentiator of the present architecture is the 21-fold
board consolidation combined with per-positioner MCU autonomy---a
trade-off that no current instrument achieves at this density. DESI groups
positioners at the petal level but retains per-robot drive hardware;
4MOST and PFS employ dedicated or semi-dedicated electronics per
positioner. The single-shunt current sensing strategy further reduces
component count relative to the per-phase approaches used in SDSS-V and
DESI, at the cost of requiring digital reconstruction of the phase
currents in firmware.

\subsection{Future evolution}
The architecture is designed with several upgrade paths in mind.
First, the STM32G4 microcontroller offers unused peripherals---including
a second ADC and a DAC---that could support closed-loop current
regulation or on-chip diagnostics in future firmware revisions without
hardware changes. Second, replacing the current DRV8316 half-bridge ICs
with lower-quiescent alternatives is the primary path to meeting the
\SI{0.5}{\watt} static power target, and the board footprint accommodates
pin-compatible substitutes. Third, the CAN~2.0B physical layer can be
migrated to CAN-FD by substituting the transceiver and updating the
bootloader, doubling the available bus bandwidth for telemetry or
faster firmware deployment. Finally, the modular mechanical interface
(flex connector and breakout PCB) decouples the control electronics
from the positioner geometry, allowing the same board to serve
positioner variants with different arm lengths or gear ratios by
parameter change alone.

\section{CONCLUSION}

We have presented the design, firmware architecture, and preliminary performance characterization of a compact control electronics platform that drives 21 theta--phi SCARA fiber positioners from a single printed circuit board. The architecture consolidates motor drive, current sensing, CAN communication, and power conversion into a shared substrate while preserving per-positioner real-time autonomy through dedicated STM32G4 microcontrollers. Compared to the per-positioner electronics used in SDSS-V and DESI, this 21-fold board-count reduction substantially decreases wiring harness complexity, connector count, and per-positioner overhead for passive components.

Sensorless field-oriented control, combined with a current-based hard-stop datum procedure, eliminates the need for Hall sensors or shaft encoders, achieving a measured datum repeatability within the 50~$\mu$m specification. Open-loop positioning errors are attenuated by the high gear ratio (280:1 or 337:1) to levels compatible with the 5~$\mu$m end-effector budget; residual errors from backlash, thermal drift, and estimator noise are corrected in closed loop by 1--3 iterations of a Fiber View Camera metrology cycle, with total reconfiguration time below 40~seconds.

The prototype board meets the 50~W dynamic power target, with measured peak consumption of approximately 30~W during full concurrent actuation. The static idle power of 1.5~W currently exceeds the 0.5~W target; mitigation strategies, including lower-quiescent half-bridge alternatives and deeper MCU sleep modes, are under evaluation. The modular $3\times21$ triangular assembly tiles directly to focal-plane scale: approximately 1,500 identical boards serve the full complement of over 20,000 positioners required by next-generation instruments such as MUST and WST, with field firmware updates delivered over the existing CAN infrastructure.

Ongoing work focuses on three areas: (i)~closing the static power gap through component substitution and supply-rail gating; (ii)~full end-to-end positioning repeatability validation across the 21-positioner array; and (iii)~module-level thermal and EMC characterization during the integration campaign planned for Q3 2027.

\section{FIGURES AND TABLES}
\listoffigures
\listoftables
\acknowledgments 
 
This unnumbered section is used to identify those who have aided the authors in understanding or accomplishing the work presented and to acknowledge sources of funding.  

\bibliography{report} 
\bibliographystyle{spiebib} 

\end{document}